\documentstyle[twoside, color, epsf]{article}
\input ibvs2.sty

\begin{document}

\IBVShead{6111}{28 June 2014}

\IBVStitletl{The 2014 Eclipse of EE\,C\lowercase{ep}: Announcement }{ for a Third International Observational Campaign}

\IBVSauth{Ga{\l}an, C.$^1$; Wychudzki, P.$^2$; Miko{\l}ajewski, M.$^2$; Tomov, T.$^2$; Dimitrov, D.$^3$}

\IBVSinst{Nicolaus Copernicus Astronomical Center, Bartycka 18, PL-00-716 Warsaw, Poland;\\
e-mail: cgalan@astri.uni.torun.pl (CG)}
\IBVSinst{Nicolaus Copernicus University, Centre for Astronomy, Gagarina 11, PL-87-100 Toru\'n, Poland;\\
e-mail: adyrbyh@gmail.com (PW); mamiko@astri.uni.torun.pl (MM); tomtom@astri.uni.torun.pl (TT)}
\IBVSinst{Institute of Astronomy and NAO, Bulg. Acad. Sc., 72 Tsarigradsko Chaussee Blvd., 1784 Sofia, Bulgaria\\
e-mail: dinko@astro.bas.bg (DD)}

\SIMBADobjAlias{EE Cep}{BD+55 2693}
\IBVStyp{   E   }
\IBVSkey{eclipsing binary, observational campaign, photometry, spectroscopy}

\IBVSabs{EE Cep is a unique system in which a Be star is eclipsed by a dark}
\IBVSabs{dusty disk, making this star similar to the famous epsilon Aur }
\IBVSabs{in many respects.  The depth and the duration of the EE Cep eclipses }
\IBVSabs{change to a large extent.  The last two eclipses were observed in the }
\IBVSabs{framework of extensive international campaigns.  The joint analysis }
\IBVSabs{of these campaigns data and historical photometry, enabled us to }
\IBVSabs{propose a model of this system, which implies a disk precession with }
\IBVSabs{a period approximately 11-12 times larger than the orbital period.  }
\IBVSabs{This model predicts that the forthcoming eclipse should be among the }
\IBVSabs{deepest observed, reaching about 2 mag.  The next eclipse approaches }
\IBVSabs{- the photometric minimum should occur around August 23, 2014.  Here }
\IBVSabs{we would like to announce a new, third international campaign with }
\IBVSabs{purpose to verify the disk precession model and to put more constraints }
\IBVSabs{on the physical parameters of this system.}

\begintext

\section{Introduction}

The variability of the 11th magnitude star EE~Cep was discovered in 1952
(E\,=\,0) by Romano (1956) and soon confirmed by Weber (1956), who reported
observations obtained during a previous eclipse in 1947 (E\,=\,$-$1). 
Thereafter all consecutive eclipses were observed with an orbital period of
5.6 yr.  The eclipses vary in an unusual way changing their shape, duration
and depth across a wide range of about 0.5 to 2.0 magnitudes (Fig.\,1)
and simultaneously with very small color variations.

\IBVSfig{6.5cm}{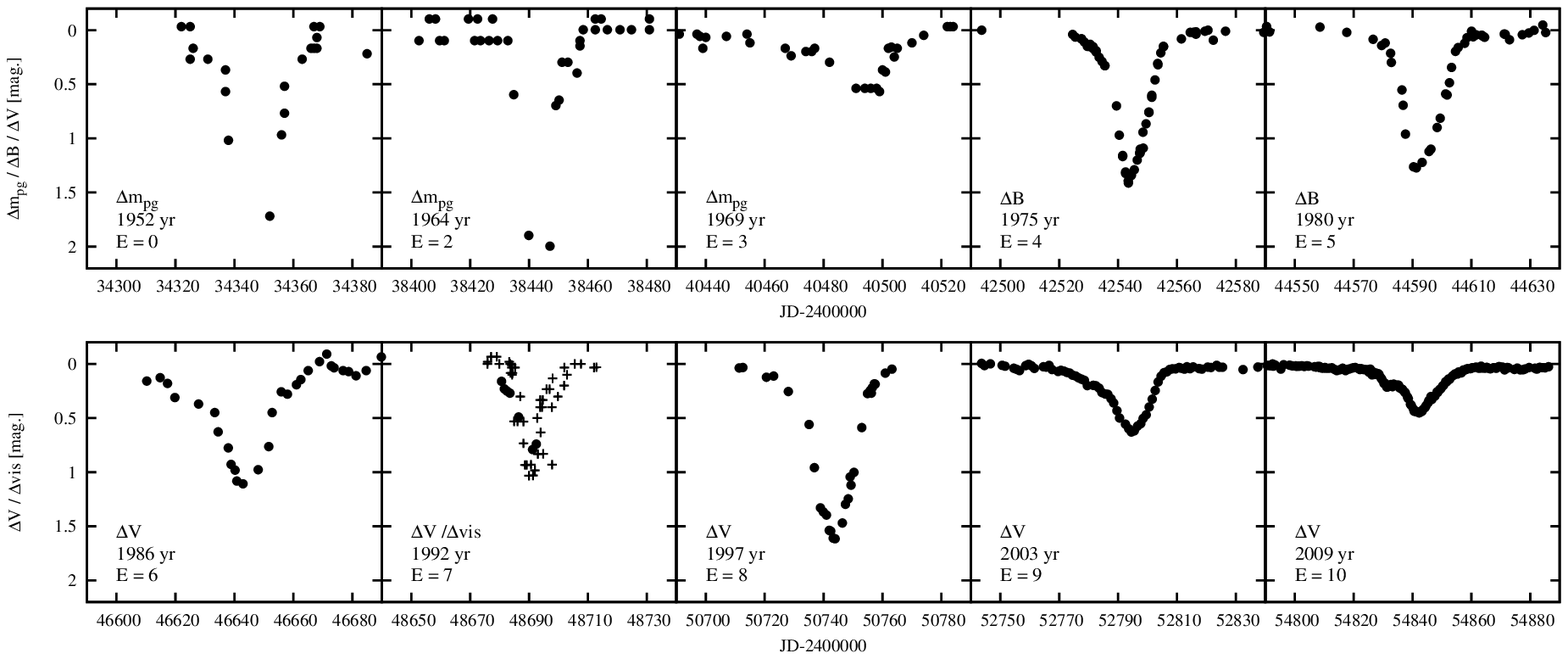}{The light curves of ten among the eleven
eclipses of EE~Cep observed since 1952 are shown.  The mean values of
Halbach's (1992) visual observations of the 1992 eclipse are marked with
crosses.}
\label{light_curve}
\IBVSfigKey{6111-f1.eps}{EE Cep}{light curve}

To explain this unusual behavior Mikolajewski~$\&$~Graczyk (1999) suggested
that the eclipses of the B5\,III primary are caused by an invisible
secondary component that consists of a dark, opaque, relatively thick disk
around a low luminosity central object: a low-mass single star or a close
binary.  In such a model, the differences between the particular eclipses
could be explained by precession, which changes the inclination of the disk
to the line of sight and the tilt of its cross-section to the direction of
motion.  Most of the eclipses have a similar, asymmetrical shape, in which
it is possible to distinguish repeatable phases of atmospheric ingress
followed by a real ingress, sloped-bottom transit during the central part of
the eclipse, and real egress followed by atmospheric eclipse (see for
details Fig.\,2 in Ga{\l}an\,et\,al. 2012).  The unique flat bottom
eclipse, observed in 1969 (E\,=\,3), can be explained by a nearly edge-on
and non-tilted projection of the disk.

The last two eclipses, in 2003 (E\,=\,9) and 2008/9 (E\,=\,10) were studied
in the framework of wide international campaigns (Miko{\l}ajewski et
al. 2005ab; Ga{\l}an et al. 2010).  The results of these campaigns
complemented by the historical light curves (Graczyk et al. 2003), enabled
us to create a model of the eclipses.  According to this model, the eclipses
are caused by a dark, geometrically thin disk precessing with period
$P_{\rm{prec}} \approx 11-12 P_{\rm{orb}}$ (Ga{\l}an et al. 2012).  The
model is based on the observations, obtained in an interval of time, almost
exactly equal to the predicted precession period.  Additional photometric
and spectroscopic observations are needed for the model verification.

\section{EE~Cep -- still a unique system}

EE~Cep is a member of the very rare class of binary systems in which the
eclipses are caused by a dark, dusty disk surrounding the orbiting companion. 
The precursor of this group is the extremely long-period ($\sim$27 yr)
$\varepsilon$\,Aur (see Guinan \& Dewarf 2002), extensively studied during
its last 2009--2011 eclipse, by the use of photometric, spectroscopic,
interferometric, astrometric, and polarimetric observations (Stencel 2013
and references therein).  Our photometric and spectroscopic observations
are published in Tomov et al. (2012) and I{\l}kiewiecz et al. (2013). 
During the last eclipse, the disk in $\varepsilon$\,Aurigae, was revealed
directly by infrared interferometric imaging (Kloppenborg et al. 2010) for
the first time.  The mechanisms of disk formation in EE~Cep and
$\varepsilon$\,Aur seems to be radically different.  Containing a B5-type
star EE~Cep has to be a very young system while $\varepsilon$\,Aur with an
F-type supergiant is significantly evolutionarily advanced.  The
observations of the eclipses provided indications for a complex structure
formed in the $\varepsilon$\,Aur (Ferluga 1990) and EE~Cep (Ga{\l}an et
al. 2008) disks.  The true nature of these structures is not known.  In the
case of $\varepsilon$\,Aur a multi-ring structure (Ferluga 1990, Leadbeater
\& Stencel 2010) was suggested but other corotating inhomogeneities cannot
be excluded (Harmanec et al. 2013).  The light curve and color variations
observed during the last two EE~Cep eclipses we interpreted in terms of a
multi-ring structure too and speculated that possible planets could be
responsible for their formation (Ga{\l}an et al. 2010).

For a long time $\varepsilon$\,Aur and EE~Cep remained the only two known
systems with dark, dusty disks as obscuring objects.  Recently however, new
cases of similar systems, with circumstellar or circumbinary disks
responsible for obscurations, emerged:\\ 1SWASP\,J140747.93−394542.6
(Mamajek et al. 2012), OGLE-LMC-ECL-11893 (Dong et al. 2014),
OGLE-LMC-ECL-17782 (Graczyk et al. 2011), M2-29 (Hajduk et al. 2008),
KH\,15D (Winn et al. 2006, Herbst et al. 2008).  This opens perspectives
for studying the dusty disk phenomenon in binary and/or multiple systems and
can be helpful to understand the origin of such disks, how they form and
evolve in various environments and on various time scales.  EE~Cep still
remains a unique case of great importance among this small sample, because
of its well-documented disk precession history during one, approximately
full, precession period (see for details Ga{\l}an et al. 2012).

\section{Call for observations}

The next eclipse (E\,=\,11) approaches and we announce a third observational
campaign.  According to the ephemeris by Mikolajewski \& Graczyk (1999) the
minimum should take place on Aug 23, 2014 $(JD_{\rm mid-ecl} = 2456893.44)$. 
Based on our model of the disk precession (Ga{\l}an et al. 2012) we can
predict that it should belong to the deepest ones reaching about 2 mag (from
10.8 mag outside of eclipse to $\sim$13 mag during the minimum in $V$
photometric band).  The longest duration eclipses observed so far occurred
in 1969 ($\sim$60 days), and 2003 and 2008/9 ($\sim$90 days).  So, we
propose to conduct systematic photometric monitoring in at least three
months time interval (July, August, September) centred on the mid-eclipse
moment.  During the previous two eclipses the blue maxima in the colors were
observed about nine days before and after the mid-eclipse.  Therefore,
special attention should be paid on precise measurements, covering about one
week time intervals around $\sim$JD 2456884 (Aug 14) and $\sim$JD 2456902
(Sep 1).  However, these moments are subject to change slightly due to
changes in the orientation of the disk.

\IBVSfig{8.0cm}{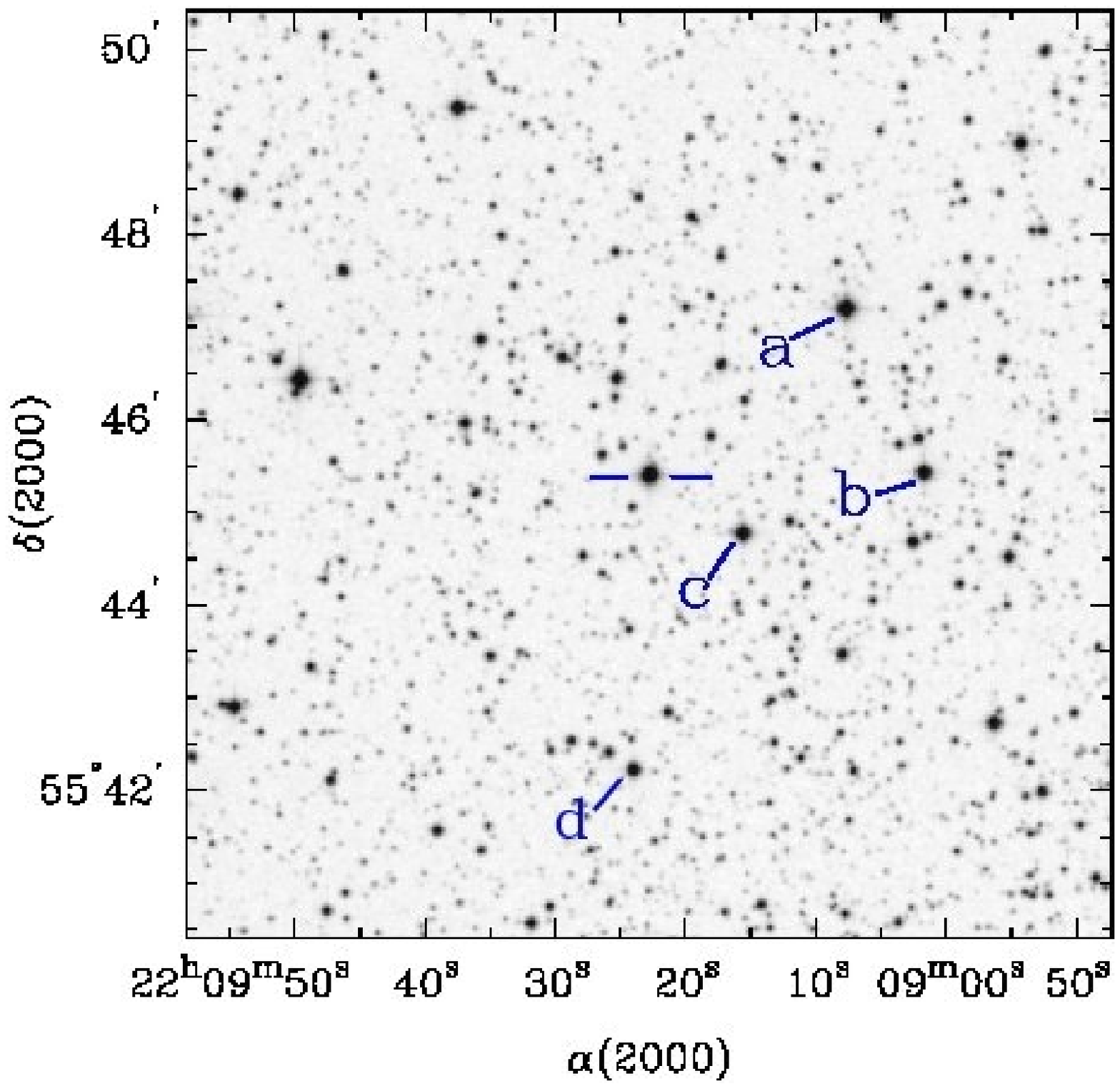}{The sky area ($10^\prime \times 10^\prime$)
around EE~Cep (reprinted from Mikolajewski et al. 2003).  The blue color
marks the comparison and the check stars recommended for the CCD
photometry.}
\label{finding_chart}
\IBVSfigKey{6111-f2.eps}{EE Cep}{finding chart}

\noindent We recommend photometric observations in the standard
Johnson-Cousins $UBV(RI)_{\mathrm C}$ system.  At least one measurement per
night with an accuracy possibly close to $\sim0\fmm01$ or better is needed. 
Some multicolour observations, far from the eclipse, should be obtained in
order to calibrate systematic differences between the observatories.  We
propose to use the four brightest objects from the Meinunger's (1975)
sequence as comparison stars.  This sequence recommended by Mikolajewski et
al. (2003) has been used in the observational campaigns during the recent
eclipses.  These stars are very close in the sky, within $\sim3^\prime$,
around EE~Cep.  In the finding chart shown in Figure\,\ref{finding_chart}
the sequence stars are marked with Meinunger's designations: ``$a$", ``$b$",
``$c$", and ``$d$" for BD+55$^{\circ}$2690, GSC-3973\,2150,
BD+55$^{\circ}$2691, and GSC-3973\,1261, respectively.  Stars ``$b$" and
``$c$" are designated as New Suspected Variables in the General Catalog of
Variable Stars (Samus et al.\,2009) -- NSV 25842, and NSV 25843,
respectively.  Baldinelli $\&$ Ghedini (1976) were the first who noted
$\sim0\fmm5$ amplitude variations of star ``$c$" based on photographic
photometry.  But they stressed the absolute necessity to confirm this result
by photoelectric method.  Miko{\l}ajewski et al. (2003) used an one channel
diaphragm photometer with a cooled photomultiplier to observe the stars
``$a$", ``$b$", and ``$c$" during the EE~Cep eclipse in 2003 and found no
significant light variations in these stars.  Star ``$b$" was suggested to be
variable on the basis of photographic photometry by Baldinelli $\&$ Ghedini
(1977), Baldinelli et al. (1981).  However, these variations were not
discussed and there was no light curve presented.

\IBVSfig{6cm}{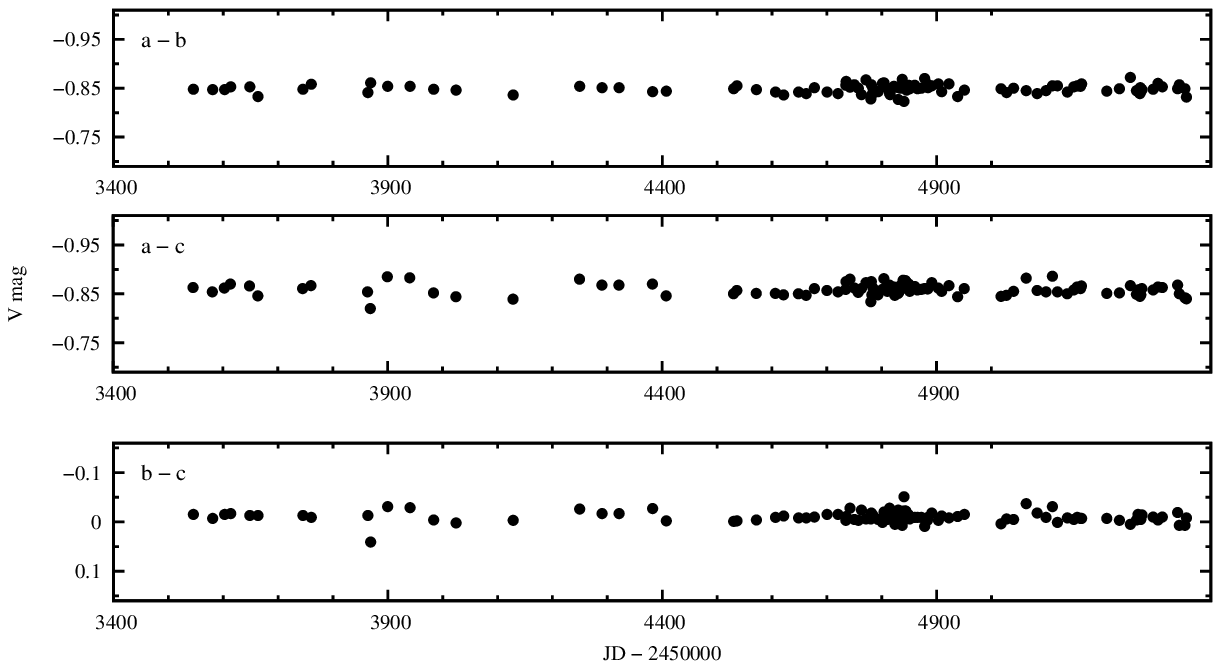}{Differential $V$ magnitudes ($a-b$, $a-c$,
and $b-c$) of the stars BD+55$^{\circ}$2690 ($a$), GSC-3973\,2150 ($b$), and
BD+55$^{\circ}$2691 ($c$) obtained in $\sim$5 years time interval during and
around the 2008/9 eclipse.}
\label{light_curve1}
\IBVSfigKey{6111-f3.eps}{EE Cep}{light curve}

We performed $UBV(RI)_{\rm C}$ CCD photometry of these stars during the
last eclipse in 2008/9.  The CCD photometry is less weather dependent than
the one channel diaphragm photometry and thus offers better accuracy for the
$BV(RI)_{\rm C}$ bands.  The differential $V$ magnitudes for stars ``$a$",
``$b$" and ``$c$" obtained during and around the last eclipse are shown in
Figure\,\ref{light_curve1}.  The observed light variations in $a-b$
differential magnitudes are insignificant and the brightness of the stars
``$a$" and ``$b$" can be recognized as a constant within the accuracy of our
photometry, which in the photometric bands $BV(RI)_{\rm C}$ is typically
$\sim0\fmm01-0\fmm02$ (depending on weather conditions during the
observations).  The differential magnitudes with respect of star “$c$” ($a-c$
and $b-c$) demonstrate somewhat larger scatter, in which small changes with
a similar pattern can be seen. Small variations of star ``$c$" with an
amplitude of a few hundredths of a magnitude cannot be excluded.  Therefore,
we recommend to use ``$a$" and ``$b$" as comparison stars and ``$c$" and ``$d$"
as check stars.

Any optical and infrared photometric as well as spectroscopic observations
obtained before, during and after the EE~Cep eclipse could turn out to be
very important.  They could make it possible to detect the mysterious
companion (disk and/or its central star/stars) in the EE~Cep system.  During
the last three orbital epochs we observed about $0\fmm2$ variations in the
$I$ passband, before and after the eclipses, which may prove a significant
contribution of a dark body in this band.  In the $JHK$ passbands, the cool
component can dominate the observed fluxes.  Moreover, these variations can
reflect in some way the changes in the spatial orientation of the disk.

\IBVSfig{6cm}{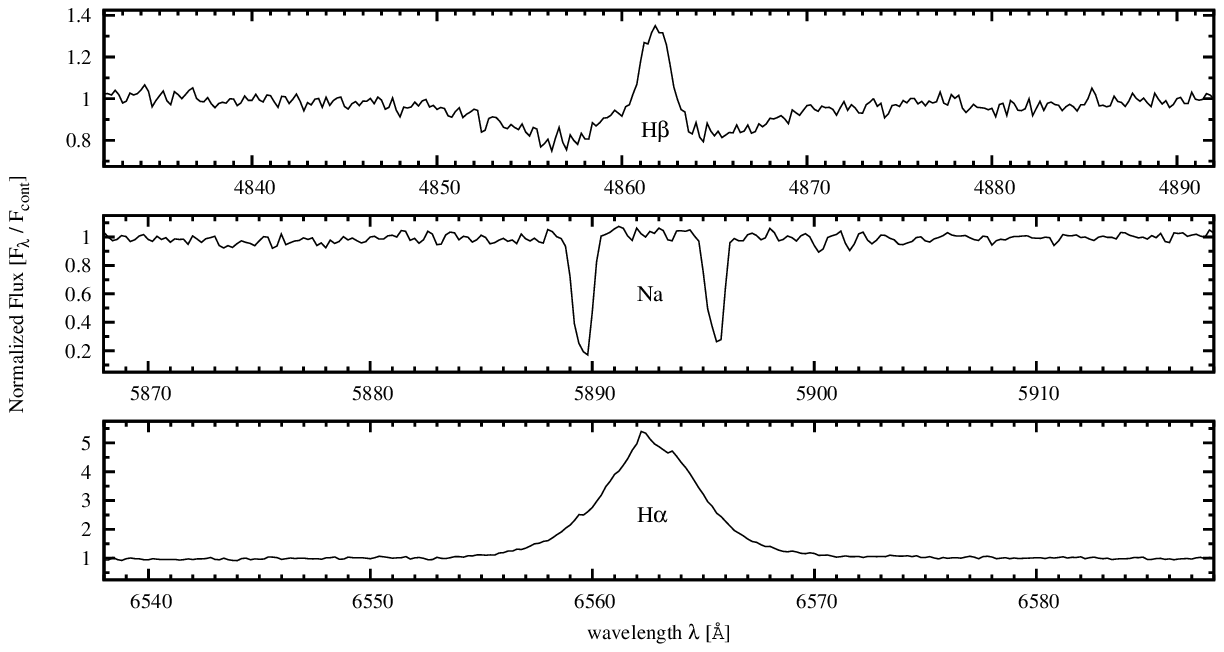}{Spectra of EE~Cep obtained on April\,8, 2014 in
the H$\alpha$, H$\beta$ and Na\,I regions (R $\sim$ 16000) using the Coud\'e
spectrograph on the 2m Ritchey--Chr\'etien telescope at Rozhen Observatory.}
\label{other1}
\IBVSfigKey{6111-f4.eps}{EE Cep}{other}

The deep EE~Cep eclipses have never been well covered with spectroscopic
observations.  Thus, it will be important to carry out systematic, high and
low resolution spectroscopic monitoring of the star during the forthcoming
eclipse.  Our spectra obtained on April\,8, 2014 in the regions of
H$\alpha$, H$\beta$, and Na\,I doublet lines (Fig.~\ref{other1}) do not
show changes caused by eclipse in the profiles of these lines (compare with
the profiles in Miko{\l}ajewski et al.  2005b).  Spectroscopic observations
obtained during the two previous eclipses in 2003 and 2008/9 show, that the
absorption lines, caused by the circumstellar matter were visible up to
about 2.5 -- 3 months before and after the mid-eclipse.  Thus, it is
advisable to cover with spectroscopic observations the period from May to
November 2014, with increased frequency of observations during the
photometric eclipse (July -- September), when we can expect significant
night to night changes in the profiles of the absorption and the emission
lines.  At least $S/N\sim30$ is recommended.  In the case of low (R$\sim$1000)
resolution observations, it would be advisable to focus mainly on the Balmer
lines evolution.  Observations of spectrophotometric standard stars are
encouraged, because they will permit us to reduce the spectra in fluxes and
to study the spectral energy distribution changes during the eclipse.

The observers can use a special web page, prepared to support the campaign
coordination {\tt http://sites.google.com/site/eecep2014campaign/}.  All
interested to participate in collecting of photometric, spectroscopic or any
other observations of the forthcoming event are encouraged to contact Piotr
Wychudzki at {\tt adyrbyh@gmail.com}.\\ 

\noindent {\bf Acknowledgements:} This study is partly supported by the
Polish National Science Centre grant No DEC-2013/08/S/ST9/00581.  This paper
is partly a result of the exchange and joint research project Spectral and
photometric studies of variable stars between Polish and Bulgarian Academies
of Sciences.  We gratefully acknowledge A.  Smith for careful reading of the
manuscript.\\

\references

Baldinelli, L., $\&$ Ghedini, S., 1976, {\it IBVS}, No. 1225 

Baldinelli, L., $\&$ Ghedini, S., 1977, {\it MmSAI}, {\bf 48}, 91

Baldinelli, L., Ferri, A., Ghedini, S., 1981, {\it IBVS}, No. 1939

Dong, S., Katz, B., Prieto, J. L., et al. 2014, {\it arXiv:1401.1195v1}

Ferluga, S., 1990, {\it A\&A}, {\bf 238}, 278 

Ga{\l}an, C., Miko{\l}ajewski, M., Tomov, T., Cika{\l}a, M., 2008, {\it IBVS}, No. 5866

Ga{\l}an, C., Miko{\l}ajewski, M., Tomov, T., et al. 2010, {\it ASPC}, {\bf 435}, 423

Ga{\l}an, C., Miko{\l}ajewski, M., Tomov, T., et al. 2012, {\it A\&A}, {\bf 544}, A53

Graczyk, D., Mikolajewski, M., Tomov, T., et al. 2003, {\it A$\&$A}, {\bf 403}, 1089

Graczyk, D., Soszy\'nski, I, Poleski, R., et al. 2011, {\it Acta Astronomica}, {\bf 61}, 103

Guinan, E. F. \& Dewarf, L. E., 2002, {\it ASP Conf. Ser.}, {\bf 279}, 121

Hajduk, M., Zijlstra, A. A., G\c{e}sicki, K., 2008, {\it A\&A}, {\bf 490}, 7

Halbach, E. A., 1992, {\it JAAVSO}, {\bf 21}, 129 

Harmanec, P., Bo{\v z}i\'c, H., Kor{\v c}\'akov\'a, D., et al. 2013, {\it CEAB}, {\bf 37}, 99

Herbst, W., Hamilton, C. M., LeDuc, K., et al. 2008, {\it Nature}, {\bf 452}, 194

I{\l}kiewicz, K., Wychudzki, P., Ga{\l}an, C., et al. 2013, {\it AASP}, {\bf 3}, 23 

Kloppenborg, B., Stencel, R., Monnier, J. D., et al. 2010, {\it Nature}, {\bf 464}, 870

Leadbeater, R., \& Stencel, R., 2010, {\it arXiv:1003.3617}

Mamajek, E. E., Quillen, A. C., Pecaut, M. J., et al. 2012, {\it AJ}, {\bf 143}, 72

Meinunger, L., 1975, {\it IBVS}, No. 965 

Mikolajewski, M. \& Graczyk, D., 1999, {\it MNRAS}, {\bf 303}, 521 

Mikolajewski, M., Tomov, T., Graczyk, D., et al. 2003,  {\it IBVS}, No. 5412 

Miko{\l}ajewski, M., Ga{\l}an, C., Gazeas, K., et al. 2005a, {\it Ap$\&$SS}, {\bf 296}, 445 \linebreak ({\tt http://www.springerlink.com/content/v6t4630310j26300/fulltext.pdf})

Miko{\l}ajewski, M., Tomov, T., Hajduk, M., et al. 2005b, {\it Ap\&SS}, {\bf 296}, 451 \linebreak({\tt http://www.springerlink.com/content/w28t429p3446p615/fulltext.pdf})

Romano, G. 1956, {\it Coelum}, {\bf 24}, 135

Samus, N.N., Durlevich O.V., Goranskij V.P., et al. 2009, {\it General Catalogue of Variable Stars (Samus+ 2007-2013)}

Stencel, R. E., 2013, {\it CEAB}, {\bf 37}, 85

Tomov, T., Wychudzki, P., Miko{\l}ajewski, M., et al. 2012, {\it BlgAJ}, {\bf 18}, a3

Weber, R. 1956, {\it Doc. des Obs. Circ.}, {\bf No. 9}

Winn, J. N., Hamilton, C. M., Herbst, W. J., et al. 2006, {\it ApJ}, {\bf 644}, 510

\endreferences

\end{document}